\RequirePackage{ifpdf}
\RequirePackage{color}
\documentclass{JHEP3}

\usepackage{amsmath}
\usepackage{amssymb}
\usepackage{cancel}
\usepackage{slashed}
\usepackage{graphicx} 

\newcommand{\bea}{\begin{eqnarray}}
\newcommand{\eea}{\end{eqnarray}}
\def\tabspace{\vrule height 3.5ex depth 2.2ex width 0pt}

\def\nl{\nonumber \\}
\def\wt{\widetilde}
\def\d{\partial}
\def\n{\nabla}
\def\l{\left(}
\def\r{\right)}
\def\lb{\left[}
\def\rb{\right]}
\newcommand{\TO}[1] {\langle  \mathbf {T} \left \{#1\right \} \rangle}
\newcommand{\Tr}[1] {{\rm{Tr}} \left \{#1\right \} }

\def\LL {\mathcal {L}}
\def\AA {\mathcal A}
\def\OO {\mathcal O}
\def\WW {\mathcal W}
\def\KK {\mathcal K}
\def\XX {\mathcal X}
\def\lc{\epsilon^{\mu\nu\rho\sigma}}

\title{The local RG equation and chiral anomalies}

\author{
Boaz Keren-Zur${}^1$ 
\\
${}^1$Institut de Th\'eorie des Ph\'enom\`enes Physiques, EPFL,\\
CH-1015 Lausanne, Switzerland\\
E-mail: 
 \email{boaz.kerenzur@epfl.ch}
}

\abstract{
We generalize the local renormalization group (RG) equation to theories with chiral anomalies.
We find that a new anomaly is required by the Wess-Zumino consistency conditions. 
Taking into account the new anomaly, the trace of the energy momentum tensor is expressed in terms of the covariant flavor currents, instead of the consistent ones. 
This result is used to show that a flavor rotation induced by the RG flow can be eliminated by a choice of scheme even in the presence of chiral anomalies.
As part of a general discussion of chiral anomalies in the presence of background sources, we also derive non-renormalization theorems.
Finally, we introduce the $\theta$ parameter as a source, and derive constraints on a perturbative running of this parameter.
}


\begin{document}

 \section{Introduction}
\label{sec_intro}

One of the main challenges in the study of quantum field theory is the characterization of 
renormalization group (RG) flows. In a sense, RG flows can be understood in terms of an approximate scale symmetry of the theory, broken explicitly by the mass parameters and the dependence on the renormalization scale.
This symmetry can be formally restored by promoting the coupling constants to background fields, and assigning them with appropriate transformation properties which compensate the non-invariance of the theory. 
The Callan-Symanzik equation is an implementation of this approach (or, at least, this is one possible interpretation of this equation), in which the transformation properties of the compensator fields under global rescaling are determined by the $\beta$ functions and anomalous dimensions.

In the framework of the local RG equation this approach is taken one step further, and the symmetry is promoted to a local one. 
The local scale transformation is realized using a background metric, thus replacing the Callan-Symanzik "symmetry" with a generalized form of the Weyl symmetry.
This methodology was first introduced by Drummond and Shore \cite{Drummond:1977dg} and was later generalized by Osborn in \cite{Osborn:1991gm} (and recently revisited in \cite{Jack:2013sha}). 
The local RG equation, which is nothing but the Ward identity associated with the Weyl symmetry, proved useful in exposing some non-trivial properties of the RG flow. The most prominent example of which is 
the perturbative proof for the gradient flow formula and the irreversibility of the flow in 4d unitary theories. This proof was obtained by invoking the Wess-Zumino consistency conditions \cite{Wess:1971yu} associated with the anomaly of the generalized Weyl symmetry.

Another approach for the study of RG flows was introduced in \cite{Komargodski:2011vj}, where a certain combination of correlation functions of the trace of the energy-momentum tensor $T$, also known as "on-shell dilaton scattering amplitude", was used to derive highly non-trivial results. Positivity constraints on a dispersion relation defined for this amplitude gave a non-perturbative proof for the $a$-theorem -- the irreversibility of the RG flow between conformal fixed points. The connection between this approach and the local RG equation was discussed in \cite{Luty:2012ww} and studied in detail in \cite{Baume:2014rla}. It was shown that, to a certain extent, the $a$-theorem and the gradient flow formula rely on the same properties of the field theory.
However, while the dilaton scattering amplitude method provides us with model independent, non-perturbative, results, 
it was demonstrated how the local RG equation can be used to translate these conclusions into constraints on the renormalization of the composite operators in the theory. These constraints were then used to prove that the asymptotic limits of perturbative RG flows are necessarily conformal fixed points.

One of the interesting points in the analysis of the local RG equation is the interplay between the generalized Weyl symmetry and global flavor symmetries. As will be explained below, the local RG equation is ambiguous in the sense that it is possible to factor out a Ward identity for these flavor symmetries and eliminate it from the equation.  
This feature corresponds to a scheme dependent artifact, in which the RG flow consists of a flavor rotation induced by wavefunction renormalization. 
An extreme example of this artifact was demonstrated in \cite{Fortin:2012hn}, where conformal theories with RG induced flavor rotations, or cyclic flows, were found.

So far, the analysis of flavor symmetries in the context of the local RG equation was restricted to anomaly free theories. In fact, the whole local RG equation formalism, as introduced in \cite{Osborn:1991gm}, was constructed under the assumption that the theory respects parity. The goal of this paper is to construct a consistent framework which allows to write the local RG equation in the presence of anomalous symmetries.

We begin with a general discussion of the compensator, or background source, method. In section \ref{sec_sources_and_flavor} we show how Ward identities for flavor symmetries are generated in this framework, and how the properties of chiral anomalies can be studied in a model independent way. We demonstrate how this method can be used to prove the Adler-Bardeen theorem \cite{Adler:1969er} for non-abelian anomalies without making any explicit computation.
As part of the exposition of chiral anomalies in the presence of background sources, we review the topic of consistent and covariant currents \cite{Bardeen:1984pm}. This terminology is relevant for the next sections, where we construct the consistent Weyl anomaly. Finally, we show how a background $\theta$ field can be used as a compensator for anomalies with dynamical gauge fields.

In section \ref{sec_local_CS} we briefly review the local RG equation framework, and the constraints imposed by the consistency with anomaly free flavor symmetries. A new contribution presented in this section is the discussion of the parity violating Weyl anomalies and the associated consistency conditions. 

The main results of the paper appear in section \ref{sec_consistency_Weyl_chiral}, where we discuss the consistency of Weyl and chiral anomalies.
We show that (a) the scheme dependent flavor rotation induced by the RG flow, mentioned above, can be consistently factorized out of the RG equation even in the presence of chiral anomalies,  (b) there are new Weyl anomaly terms that must be added to the local RG equation in order to satisfy the Wess-Zumino consistency condition. These new terms have a simple interpretation - when using the local RG equation to express $T$ in terms of the composite operators, one has to use the covariant flavor current instead of the consistent one. (c) We use the gradient flow equation to find a formula which relates between the (perturbative) running of the $\theta$ parameter and the remaining $\beta$ functions in the theory.

 \section{The background source method and chiral anomalies}
\label{sec_sources_and_flavor}
In this section we discuss the background source method and how it can be used to
analyze the global symmetries of the system. We show how to write anomalous Ward identities in this framework, and how 
to constrain the structure of the anomalies. We also introduce a $\theta(x)$ background field as a compensator for anomalies with
dynamical fields.

 \subsection{The generating functional}
\label{sec_generating_functional}

Consider a four dimensional conformal fixed point. In order to study correlation functions of the energy momentum tensor $T_{\mu\nu}(x)$ and other composite operators in the theory $\OO_I(x)$, we introduce a background metric $g^{\mu\nu}(x)$ and background sources $\lambda^I(x)$, and define the effective action as
\bea
\label{eq_WW_g_lambda}
 e^{i {\cal W}[g,\lambda]}\equiv \int {\cal {D}} \Phi e^{i S[\Phi, g]+i\int d^4x \sqrt {-g}\lambda \OO}~.
\eea
The functional $\WW[g,\lambda]$ is assumed to be renormalized (without specifying the particular regulator or renormalization procedure), such that 
the derivatives of $\WW$ with respect to the metric and the $\lambda$ sources generate the renormalized time-ordered correlation functions of the composite operators in the theory
\bea
\label{eq_T_correlator}
\frac{-i}{\sqrt{ - g}}\frac{\delta }{\delta \lambda^I(x)} \ldots \frac{-2i}{\sqrt{ - g}}\frac{\delta }{\delta g^{\mu\nu  }(y)} i \WW [g,\lambda] \Big |_{g=\eta,\lambda=0}&=& \TO { \OO_I(x)\ldots T_{\mu\nu}(y)}~.
\eea
We will use the following notations
\bea
\label{eq_renormalized_operator}
\lb T_{\mu\nu}(x)\rb&\equiv& \frac{2}{\sqrt{ - g}}\frac{\delta }{\delta g^{\mu\nu}(x)}  \nl
 \lb \OO_I (x)\rb&\equiv&\frac{1}{\sqrt{ - g}} \frac{\delta }{\delta \lambda^I(x)}
\eea
to express the fact that a derivative with respect to the sources corresponds to an insertion of a renormalized composite operator.

The functional derivatives, evaluated in a background with vanishing background sources, corresponds to the correlation functions of the operators at the fixed point.
A non-zero background value for such a source is equivalent to the introduction of an interaction term in the Lagrangian, therefore the same framework can be used to study the dynamics of the system in the presence of small perturbations.

\subsection{The sources as compensators for flavor symmetries}
\label{sec_flavor_compensators}
The background fields act as sources for the composite operators, but they can be understood also as compensators for explicitly broken symmetries. Consider a fixed point which possess a global internal "flavor" symmetry. The operators in the spectrum must reside in representations of this flavor symmetry
\bea
\delta_\alpha \OO_I &=& \alpha ^A (T_A)_I^J\OO_J
\eea
where $T_A$ are the generators of the symmetry transformation (in our notations, these are anti-hermitean matrices satisfying $[T_A,T_B]=f^C_{AB}T_C$) and $\alpha^A$ is an arbitrary transformation parameter. The couplings $\lambda^I \OO_I$ appearing in \eqref{eq_WW_g_lambda} break this symmetry explicitly, however the symmetry can be formally restored if we assign the opposite transformation properties to the sources
\bea
\label{eq_lambda_flavor_transformation}
\delta_\alpha \lambda^I &=& -\alpha ^A (T_A)_J^I\lambda^I~.
\eea
In terms of the generating functional (where the dynamical fields are already integrated out) the existence of the symmetry corresponds to an invariance of $\WW$ under global rotations of the sources
\bea
\label{eq_global_invariance}
-\alpha^A \int d^4x (T_A\lambda)^I \frac{\delta}{\delta \lambda(x)} \WW [g,\lambda]&=&0~.
\eea 
If we assign a non-zero value to the $\lambda$'s, then $\WW$ describes a system where the flavor symmetry is explicitly broken (it also describes a theory which is perhaps no longer scale invariant, but this will be discussed in the next section). However, we can still use eq. \eqref{eq_global_invariance} and derive non-trivial constraints regarding the symmetry breaking pattern of the effective field theory (see, for example, \cite{Gasser:1983yg}).

The fact that the sources are $x$-dependent allows us to write Ward identities for these global symmetries in a convenient form. For this purpose we introduce background gauge fields $A_\mu=A_\mu^AT_A$ which act as sources for the Noether currents
\bea
\lb J_A^\mu (x)\rb &\equiv& \frac{1}{\sqrt{ - g}} \frac{\delta }{\delta A_\mu^A(x)}~.
\eea  
We also promote the transformation parameter $\alpha^A$ to be an $x$-dependent function, and assign the following transformation property to the background gauge field
\bea
\label{eq_A_flavor_transformation}
\delta_\alpha A_\mu &=& \d_\mu \alpha + [A,\alpha] \equiv \n_\mu\alpha~.
\eea 
In the absence of anomalies, eq. \eqref{eq_global_invariance} can now be promoted to
\bea
\label{eq_global_invariance2}
\Delta^F_{\alpha} \WW[g,\lambda,A]=0
\eea
where $\Delta^F_{\alpha}$ is the following non-local generator of the flavor symmetry
\bea
\Delta^F_{\alpha} &\equiv&
\int d^4x\l \n_\mu \alpha^A    \frac{\delta}{\delta A^A_\mu(x)} 
- \alpha^A (   T_A \lambda)^I  \frac{\delta}{\delta \lambda^I(x)} \r 
\eea
where we used equations \eqref{eq_lambda_flavor_transformation} and \eqref{eq_A_flavor_transformation}.
Notice that the operator defined in this way satisfies the algebra
\bea
\label{eq_flavor_algebra}
[\Delta^F_\alpha,\Delta^F_{\beta}]&=&\Delta_{[\alpha,\beta]}^F~.
\eea

Eq. \eqref{eq_global_invariance2} is the generator of Ward identities for the flavor symmetry. Indeed, by taking $\alpha^A(x)=\alpha^A \frac{\delta(x-x_0)}{\sqrt {-g}}$, and using the notation for renormalized operators introduced in \eqref{eq_renormalized_operator}, eq. \eqref{eq_global_invariance2} can be written as
\bea
\nabla_\mu \lb J_A^\mu \rb &=&- (T_A \lambda)^I \lb \OO_I\rb
\eea
In order to obtain the Ward identities for correlation functions we use eq. \eqref{eq_global_invariance2} to write
\bea
\label{eq_pre_WI}
\Delta_\alpha^F \frac{\delta}{\delta\lambda^{I_1}(x_1)}\ldots \frac{\delta}{\delta\lambda^{I_n}(x_n)}\WW 
&=&
\left [\Delta_\alpha^F ,\frac{\delta}{\delta\lambda^{I_1}(x_1)}\ldots \frac{\delta}{\delta\lambda^{I_n}(x_n)}\right]\WW 
\eea
or equivalently, by using eq. \eqref{eq_T_correlator},
\bea
&&\alpha^A\nabla_{\mu,x_0} \TO{ J_A^\mu(x_0) \OO_{I_1}(x_1)\ldots\OO_{I_n}(x_n)} 
-\TO{
\delta_\alpha  \LL(x_0) \OO_{I_1}(x_1)\ldots\OO_{I_n}(x_n)} \nl
&&~~~~~=i\sum_{i=1}^n \delta(x_0-x_i) \TO{ \OO_{I_1}(x_1)\ldots\delta_\alpha \OO_{I_i}(x_i)\ldots\OO_{I_n}(x_n) }
\eea
where $\delta_\alpha  \LL\equiv- (T_A \lambda)^I \lb \OO_I\rb $.

\subsection{Anomalies}
\label{sec_chiral_anomalies}
Anomalies appear when the quantum theory cannot be regularized in a way which preserves all the symmetries of the classical theory.
This is manifested in the appearance of scheme independent contact terms in Ward identities. 
In order to express this effect in the background source formalism, we introduce a local functional on the RHS of eq. \eqref{eq_global_invariance2} 
\bea
\label{eq_flavor_anomaly}
\Delta^F_{\alpha} \WW[g,\lambda,A]=\int d^4x\AA_\alpha^F[g,\lambda,A]~.
\eea
In a background with vanishing sources, $\AA_\alpha^F$ vanishes, and the symmetry seems to be exact. 
However, $\AA_\alpha^F$  encodes the anomalous contact terms in the Ward identity. Indeed, eq. \eqref{eq_pre_WI} can now be written as
\bea
\label{eq_flavor_anomaly_WI}
\Delta_\alpha^F \frac{\delta}{\delta\lambda^{I_1}(x_1)}\ldots \frac{\delta}{\delta\lambda^{I_n}(x_n)}\WW 
&=&
\left [\Delta_\alpha^F ,
\frac{\delta}{\delta\lambda^{I_1}(x_1)}\ldots \frac{\delta}{\delta\lambda^{I_n}(x_n)}\right]\WW \nl
&& + \frac{\delta}{\delta\lambda^{I_1}(x_1)}\ldots \frac{\delta}{\delta\lambda^{I_n}(x_n)} \int d^4x \AA_\alpha^F
\eea
and in terms of the correlation functions:
\bea
&&\alpha^A\nabla_{\mu,x_0} \TO{ J_A^\mu(x_0) \OO_{I_1}(x_1)\ldots\OO_{I_n}(x_n)} 
-\TO{
\delta_\alpha  \LL(x_0) \OO_{I_1}(x_1)\ldots\OO_{I_n}(x_n)} \nl
&&~~~~~=i\sum_{i=1}^n \delta(x_0-x_i) \TO{ \OO_{I_1}(x_1)\ldots\delta_\alpha \OO_{I_i}(x_i)\ldots\OO_{I_n}(x_n) }
\nl&& ~~~~~~~~~~~~
+ \frac{\delta}{\delta\lambda^{I_1}(x_1)}\ldots \frac{\delta}{\delta\lambda^{I_n}(x_n)} \int d^4x \AA_\alpha^F~.
\eea

The form of the anomaly $\AA_\alpha^F$ can be obtained by considering all possible Ward identities and looking for the appearance of scheme independent contact terms. An alternative approach is to write the most general function $\AA_\alpha^F$ allowed by  power counting and by the following constraints: first of all,  the anomaly function cannot consist of a term which can be written as a variation of a local functional
\bea
\int d^4x \AA_\alpha^F&\neq &\Delta_\alpha^F \int d^4x \sqrt {-g} f [g,\lambda,A] ~.
\eea
Such an anomaly can be eliminated by a redefinition of the generating functional
\bea
\WW\to \WW - \int d^4x \sqrt {-g} f [g,\lambda,A] 
\eea
or, equivalently, the existence of the contact terms depends on a choice of scheme, in contradiction to the definition of anomalies given above.
The second constraint, known as the Wess-Zumino condition \cite{Wess:1971yu}, is derived from the algebra \eqref{eq_flavor_algebra} acting on the generating functional
\bea
\label{eq_WZ_CC}
\Delta^F_\alpha\l \int d^4x \AA^F_\beta\r  - \Delta^F_{\beta}\l \int d^4x \AA^F_\alpha\r &=&\int d^4x \AA^F_{[\alpha,\beta]}~.
\eea

For non-abelian symmetries,  the combination of the two constraints is enough to pin down the exact structure of the anomaly (up to numerical normalization). In the presence of the background gauge fields $A_\mu^A$ only, it is well known (see, e.g.  \cite{Weinberg:1996kr}) that the unique solution (up to a numerical factor) is 
\bea
\label{eq_chiral_anomaly}
\AA_\alpha^F &=&  \d_\mu \alpha^A  \KK^{\mu}_A
\eea
where
\bea
\label{eq_chiral_anomaly_K}
\KK_A^\mu &=&\frac{1}{48\pi^2} \lc  \Tr {T_A A_\nu \l 2\d_\rho A_\sigma  +  A_\rho A_\sigma\r }~.
\eea
In appendix \ref {app_WZ_CC_BG_sources} we show that this is true also in the presence of the background metric and the $\lambda$ sources. Notice that there are necessarily new consistent anomaly terms in the presence of background sources. Such terms are generated when performing a redefinition of the background sources\footnote{This redefinition corresponds to a change of  scheme for the renormalized composite operators, see sec. 2.2.5 of \cite{Baume:2014rla}.}
\bea
A^A_\mu &\to& {A^A_\mu}'=A_\mu^A+f_I ^A\n_\mu\lambda^I~.
\eea
The goal of the analysis in the appendix is to verify that all possible consistent anomalies involving the background metric and $\lambda$ sources can indeed be eliminated by a choice of scheme.

One consequence of the analysis appearing in the appendix is that a $\lambda$ dependent coefficient in front of the anomaly \eqref{eq_chiral_anomaly}, namely, an anomaly of the form
\bea
 f(\lambda)\d_\mu \alpha^A  \KK^{\mu}_A
\eea
is inconsistent. 
The fact that the anomaly coefficient must be independent of the coupling constants $\lambda$, 
implies that the contributions to the anomalous diagram are exhausted at 1-loop. This is a quick derivation of the Adler-Bardeen theorem regarding the non-renormalization of the non-abelian anomaly.

Let us mention a few aspects in which the anomaly of an abelian symmetry is different from the non-abelian case.  First, the anomaly of a $U(1)$ axial symmetry can involve currents coupled to dynamical gauge fields. This scenario will be discussed in section \ref{sec_theta}. Second, in a curved background, an abelian symmetry can have an anomaly proportional to the Pontryagin density
\bea
\label{eq_chiral_gravitatinal_anomaly}
\AA_\alpha^F
&=& \alpha^Ad_A\epsilon^{\mu\nu\alpha\beta}R_{\mu\nu\rho\sigma}R_{\alpha\beta}^{~\rho\sigma}
\eea
where $d_A=-\frac{1}{384\pi^2} \Tr{T_A}$. 
The last point relevant to our discussion is that  the Wess-Zumino condition for abelian symmetries is not quite enough to fully characterize the anomaly. The constraints on the singlet anomaly are discussed in section \ref{sec_consistency_Weyl_global_chiral}.

\subsection{Consistent and covariant currents}
\label{sec_consistent_covariant}

We will now briefly review the terminology of consistent and covariant currents \cite{Bardeen:1984pm}. The reason for introducing this notation here is that the vector $\XX_A^\mu$ defined below will end up playing an important role in section \ref{sec_consistency_Weyl_global_chiral} when we present the consistent Weyl anomaly.

In \cite{Bardeen:1984pm}, the anomalous Ward identity \eqref{eq_flavor_anomaly_WI} was used to derive the transformation properties of the  currents in the presence of anomalies
\bea
\delta_\alpha^F  J_A^\mu &\equiv&
\Delta_\alpha^F \frac{\delta}{\delta A_\mu^A (x)}\WW \nl
&=&
f^C_{BA}\alpha^B \frac{\delta}{\delta A_\mu^C (x)}\WW
 + \alpha^A\frac{\delta}{\delta A_\mu^A (x)} \int d^4 y \d_\nu \alpha^B  \KK_B^{\nu}~.
\eea
The interpretation of the anomalous term on the second line is that $J_A^\mu$, the current obtained by taking a derivative with respect to the background gauge field $A_\mu^A$, also known as the \emph{consistent} current, does not transform covariantly under the flavor rotation.
The authors of \cite{Bardeen:1984pm} define a \emph{covariant} current, which is a combination of the composite operator and a function of the the background fields $ \XX^\mu_A(A)$, which transform covariantly:
\bea
\label{eq_covariant_current}
\tilde J^\mu_A &=& J^\mu_A + \XX^\mu_A(A)\nl
\delta_\alpha^F 
\tilde J_\mu^A &=&
f^C_{BA}\alpha^B\tilde J^\mu_C
\eea
This implies that $\XX_\mu^A(A)$ must be a functional of the sources which satisfies
\bea
\label{eq_transformation_XX}
\Delta_\alpha^F 
\XX_\mu^A &=&
f^C_{BA}\alpha^B\XX^\mu_C
 - \alpha^A\frac{\delta}{\delta A_\mu^A (x)} \int d^4 y \d_\nu \alpha^B  \KK_B^{\nu}~.
\eea
The solution for \eqref{eq_transformation_XX} was found to be
\bea
\XX_A^\mu&=&
\frac{1}{48\pi^2}\lc \Tr{T_A 
  A_\nu F_{\rho\sigma} +  F_{\rho\sigma}A_\nu -A_\nu A_\rho A_\sigma
}~.\eea
Notice that the covariant current satisfies a covariant conservation equation
\bea
\label{eq_conservation_covariant_current}
\n_\mu \wt J_A^\mu &=& 
\d_\mu \KK^\mu_A + \n_\mu \XX_A^\mu = d_{ABC}
 \lc F_{\mu\nu}^B F_{\rho\sigma}^C\eea
where $d_{ABC}\equiv \frac{1}{64\pi^2} \Tr {T_A \{T_B, T_C\}}$~.

\subsection{Anomalies with dynamical gauge fields and the $\theta$ parameter}
\label{sec_theta}
So far, our discussion was restricted to cases where the anomalous Ward identity involved global symmetries only.
In gauge theories the anomalous non-conservation of an axial $U(1)$ current can appear in the form of a composite operator of the dynamical gauge fields. 
In this case, the Ward identity can be written by introducing a new compensator field $\theta(x)$:
\bea
\label{eq_axial_generator}
\Delta^{5}_\alpha=
\int d^4x\l  
 \d_\mu \alpha^5    \frac{\delta}{\delta A^5_\mu(x)} - \alpha^5  (T_{5} \lambda)^I  \frac{\delta}{\delta \lambda^I(x)} 
+ \alpha^5
\frac{\delta}{\delta \theta(x)}\r 
\eea
where $A^5_\mu(x)$ is the source associated with the axial current $J^{5\mu}$ (to avoid cluttering the notations we will assume that there is only one axial $U(1)$ symmetry). $\theta$ can be understood as the source for the renormalized anomaly operator
\bea
\label{eq_theta_and_F_tildeF}
\lb\d_\mu  \hat \KK_{5}^\mu\rb
&\equiv&\frac{\delta}{\delta \theta(x)}
\eea
where $\hat \KK_{5}^\mu$ is a non-gauge invariant function of the dynamical gauge fields.
As can be read from \eqref{eq_axial_generator}, 
the source $\theta(x)$ transforms under axial rotations by shifts
\bea
\delta_{\alpha_5} \theta&=&\alpha_5 ~.
\eea

$\theta(x)$ is a source for an operator which is a total derivative, and therefore it does not contribute to perturbative computations. More precisely, there might be dependence on the gradient of $\theta(x)$, but perturbation theory is insensitive to its 
zero momentum component. In section \ref{sec_running_theta} we discuss some implications that can be derived from the fact that the $\beta$-functions and anomaly coefficients are independent of $\theta$. 

Now, the fact that the anomaly is a total derivative is not manifest in this formalism. 
In fact, since $\theta$ is a dimensionless source, it is not obvious that the renormalized anomaly operator does not mix with marginal operators, a possibility which would invalidate the above argument regarding the $\theta$-independence of the $\beta$-functions.
This difficulty can be addressed as follows:
imagine that we could assign a background gauge field $A^{\KK_5}_\mu(x)$ as a source for the operator $\hat \KK_{5}^\mu$. Such a field could be used to write the anomalous Ward identity \eqref{eq_axial_generator} as
\bea
\label{eq_axial_generator_naive}
\Delta^{5,naive}_\alpha\overset{?}{=}
\int d^4x\l 
 \d_\mu \alpha^5    \frac{\delta}{\delta A^5_\mu(x)} - \alpha^5  (T_{5} \lambda)^I  \frac{\delta}{\delta \lambda^I(x)} 
- \d_\mu \alpha^5    \frac{\delta}{\delta A^{\KK_5}_\mu(x)}\r ~.
\eea
In general, however, it is impossible to couple a background source to a gauge non-invariant operator. Nevertheless, since the divergence of $\hat \KK_{5}^\mu$ is gauge invariant, such a coupling is possible if we impose the constraint that $A_\mu^{\KK_5}$ is a gradient of a scalar function
\bea
A^{\KK_5}_\mu(x)=\d_\mu\theta
\eea
(recall that we restrict our discussion to perturbation theory, thus we can integrate by parts and ignore boundary terms and instanton effects). 
Under this constraint, eq. \eqref{eq_axial_generator_naive} is equivalent to \eqref{eq_axial_generator}. We conclude that, as claimed above, $\theta$ appears in this formalism only via its gradient.

\section{The local RG equation}
\label{sec_local_CS}

In this section we review the necessary ingredients of the local RG equation framework. For more details we refer the reader to \cite{Baume:2014rla}.

\subsection{The Weyl symmetry}
Let us now turn off the background sources $\lambda^I$ and discuss the theory in a curved background.
Since the theory is assumed to be conformal, $\WW[g]$ is invariant (up to anomalies) under the Weyl symmetry, a local rescaling of the background metric
\bea
\delta^W_\sigma g^{\mu\nu}&=&2\sigma g^{\mu\nu}~.
\eea
In order to express the associated Ward identity we follow the same procedure discussed above for flavor symmetries, namely define the symmetry generator
\bea
\Delta_\sigma^W
&=&\int d^4x  ~
2\sigma  g^{\mu\nu}(x)\frac{\delta}{\delta g^{\mu\nu}(x)}
\eea
and write the anomalous conservation equation as
\bea
\Delta_\sigma^W \WW &=& \int d^4x \AA^W_\sigma~,
\eea
where the most general expression for the scheme independent, consistent, Weyl anomaly is 
\bea
\label{eq_Weyl_anomaly}
\AA^W_\sigma&=&\sqrt {g} \l a E_4 - cW^2\r+e ~
\epsilon^{\mu\nu\alpha\beta}R_{\mu\nu\rho\sigma}R_{\alpha\beta}^{~\rho\sigma} ~.
\eea
$E_4$ is the 4 dimensional Euler density and $W^2$ is the Weyl tensor squared. $a$, $c$ and $e$ are numerical coefficients which depend on the details of the theory.
The parity violating Weyl anomaly is discussed in \cite{Nakayama:2012gu}.

\subsection{The Weyl symmetry off-criticality}
Turning on the sources $\lambda^I$, it is clear that the Weyl symmetry is broken at the classical level if the corresponding operators are not marginal.
As in the case of flavor symmetries, one can use the sources as compensators for the Weyl symmetry, by assigning them with the appropriate linear transformation
\bea
\label{eq_Delta_W_naive}
\Delta_\sigma^{W,classical}
&=&\int d^4x  ~
\sigma \l 2g^{\mu\nu}(x)\frac{\delta}{\delta g^{\mu\nu}(x)} + (4\delta_I^J-d_I^J) \lambda^I(x) \frac{\delta}{\delta \lambda^J(x)}\r 
\eea
where the Weyl weight matrix $d_I^J$ is assumed to be written in a diagonal form.
This procedure is not sufficient for a proper discussion of the Weyl symmetry in the quantum theory. 
Indeed, away from the fixed point ($\lambda^I\neq 0$), the Weyl symmetry is broken by quantum effects, and it is necessary to take into account the anomalous dimensions of the operators. This is achieved by generalizing eq. \eqref{eq_Delta_W_naive} to include the most general, non-linear, transformation allowed by dimensional analysis and symmetry constraints. 

Focusing for the moment on sources for marginal operators, the most general Weyl generator can be parameterized as
\bea
\label{eq_Weyl_generator_beta}
\Delta_\sigma^W
&=&\int d^4x \sigma 
\l 2 g^{\mu\nu}\frac{\delta}{\delta g^{\mu\nu}(x)}
-\beta^I(\lambda)\frac{\delta}{\delta \lambda^I(x)}\r 
\eea
where $\beta^I(\lambda)$ are model dependent functions. 
The anomalous conservation equation now takes the form
\bea
\Delta_\sigma^W \WW [g,\lambda] &=&\int d^4x \AA_\sigma^W[g,\lambda]~.
\eea

Before discussing the form of the anomaly, let us comment on the interpretation of this operator.
This operator can be understood as the generator of "local RG flows" in the following sense: The background value for the sources is defined in a specific renormalization scale $\mu$. A rescaling of the $\mu$ (as well as the mass parameters of the theory, but here we still consider only marginal deformations) can be compensated by a global rescaling of the metric. This can be expressed as follows
\bea
\Delta^\mu \WW \equiv   
\l  \mu\frac{\d}{\d\mu} + 2 \int d^4x g^{\mu\nu}\frac{\delta}{\delta g^{\mu\nu}(x)}\r  \WW =0~.
\eea
Using eq. \eqref{eq_Weyl_generator_beta} to eliminate the dependence on the metric we define the generator of RG transformation as
\bea
\Delta^{RG} \WW &\equiv& \l\Delta^\mu +  \Delta^W_{\sigma = -1}\r \WW\nl
  &=&\l \mu\frac{\d}{\d\mu}+\int d^4x ~\beta^I\frac{\delta}{\delta \lambda^I(x)}\r \WW=0~.
\eea
This establishes the connection between  $\Delta_\sigma^W$, the generator of the Weyl symmetry off-criticality, and the Callan-Symanzik equation.

The Ward identity for the Weyl symmetry (ignoring for the moment possible anomalies) has the following form
\bea
[T]=\beta^I[\OO_I]
\eea
where $T$ is the trace of the energy momentum tensor. For correlation functions it is given by
\bea
\label{eq_Weyl_WI}
&&\TO{\l T(x_0) -\beta^I \OO_I(x_0)
\r \OO_{I_1}(x_1)\ldots\OO_{I_n}(x_n)} \nl
&&~~~~~=i\sum_{i=1}^n \delta(x_0-x_i) \TO{ \OO_{I_1}(x_1)\ldots\d_{I_i}\beta^J \OO_{J}(x_i)\ldots\OO_{I_n}(x_n) }~.
\eea
 The matrix $\d_I\beta^J$ can thus be interpreted as the anomalous dimension matrix for the nearly-marginal operators $\OO_I$.

\subsection{The Weyl symmetry and flavor symmetries}
\label{sec_Weyl_flavor_no_anoamlies}

From this point on we will focus on fixed points perturbed only by nearly-marginal deformations, and we will use $\lambda^I(x)$ to denote dimensionless sources. Now, if the conformal fixed point possess some global symmetries, 
then the spectrum of dimension 4 contains the descendent operator $\n_\mu J^\mu_A$ where $J^\mu_A$ is the consistent current\footnote{If the theory has a dimension 2  scalar operators $\OO_a$, then one has to consider also the dimension 4 operators $\n^2 \OO_a$. The importance of these operators is discussed in detail in \cite{Baume:2014rla}.}. As discussed in section \ref{sec_flavor_compensators},
the source associated with this operator is the background gauge field $A_\mu^A$.

The most general parametrization of the generator of Weyl transformations in the presence of the dimensionless sources and the background
gauge fields can be given by
\bea
\label{eq_Delta_W}
\Delta_\sigma^W&=&\int d^4x  
\l 2 \sigma g^{\mu\nu}\frac{\delta}{\delta g^{\mu\nu}(x)}
-\sigma \beta^I\frac{\delta}{\delta \lambda^I(x)}
-\l \sigma \rho^A_I\n_\mu\lambda^I - \d_\mu\sigma S^A\r \frac{\delta }{\delta A_\mu^A(x)}
\r 
\eea
In terms of the renormalized operators, this can be written as an operator equation:
\bea
\label{eq_T_beta_S}
[T]&=&
\beta^I[\OO_I]+S^A \n_\mu [J^\mu_A]
\eea

So far, we simply parameterized the symmetry generator in terms of unknown functions basing on naive dimensional analysis alone. Let us now derive some features of this operator based on symmetry considerations. A first constraint on this general parameterization of the generator of Weyl transformation is that it has to commute with the generators of the flavor symmetries of the theory
\bea
\label{eq_Weyl_chiral_consistency} 
\left [\Delta_\sigma^W , \Delta_\alpha^F\right] = 0~.
\eea
This implies that the functions $\beta^I$, $\rho_I^A$ and $S^A$ must be covariant functions of the sources, 
and that the derivatives of the sources must be replaced by covariant derivatives, where we use the following notations:
\bea
\label{eq_covariant_derivatives}
\n_\mu \lambda^I &=& \d_\mu\lambda^I + A_\mu^A(T_A\lambda)^I\nl
\n_\mu\theta&=& \d_\mu\theta-A_\mu^5
\eea

The parameterization \eqref{eq_Delta_W} has a built in ambiguity. Indeed, if the global symmetry is not-anomalous one can consider a local RG equation constructed from a combination of Weyl transformations and flavor symmetry rotations:
\bea
\label{eq_Weyl_ambiguity}
{\Delta_\sigma^W}' \WW = \l \Delta^W_{\sigma} + \Delta^F_{\alpha=\sigma \omega}\r \WW &=&\int d^4x \AA^W_\sigma
\eea
This can be interpreted as a redefinition of the generator of Weyl transformation given by
\bea
\label{eq_operator_gauge_freedom}
{\beta^I}'&=& \beta^I+\l \omega^A T_A \lambda\r^I\nl
{S^A}'&=& S^A+\omega^A\nl
{\rho_I^A}'&=& \rho_I^A-\d_I\omega^A~.
\eea
(this is true only when the parameter $\omega^A$ is a covariant function of the source, otherwise \eqref{eq_Weyl_chiral_consistency} is violated). As discussed in \cite{Luty:2012ww}, this ambiguity can be traced back to the freedom to choose non-symmetric wavefunction renormalization.
It is possible, however, to define non-ambiguous functions
\bea
\label{eq_def_B}
B^I&=& \beta^I-\l S^AT_A\lambda\r^I\nl
 P_I^A&=&\rho_I^A +\d_I S^A~.
 \eea
By adding and subtracting the flavor Ward identity with parameter $\omega^A=-S^A$ we find that the generator of Weyl symmetry takes the form 
\bea
\label{eq_extracting_flavor_rotation}
\Delta_\sigma^W&=&\int d^4x \sigma 
\l 2 g^{\mu\nu}\frac{\delta}{\delta g^{\mu\nu}(x)}
-B^I\frac{\delta}{\delta \lambda^I(x)}
- B_\mu^A \frac{\delta }{\delta A_\mu^A(x)}
\r -\Delta^{F}_{\sigma S}
\eea
where the vector beta function is given by $B_\mu^A = P_I^A\n_\mu\lambda^I$ (for a detailed discussion of this function, see \cite{Nakayama:2013ssa}). 
Using this notation we see that the $S^A$ dependent part of the generator describes a flavor rotation induced by the RG flow, which can be eliminated by a choice of renormalization scheme.

For completeness, we mention the last constraint on the form of the generator of the Weyl symmetry, which is derived from the algebra of the Weyl symmetry
\bea
\label{eq_Weyl_algebra}
\left [ \Delta^W_{\sigma'},\Delta_{\sigma}^W\right]&=&0~.
\eea
This consistency condition implies the following constraint on the functions 
\bea
\label {eq_BP}
B^I P_I^A&=&0~.
\eea
The implications of this constraint are discussed in \cite{Baume:2014rla}, and will not play a crucial role in this work.

 \subsection{The Weyl anomaly}
\label{sec_local_CS_anomaly}

As in the case of the chiral anomalies, the Weyl anomaly $\AA_\sigma^W$ encodes contact terms in the Weyl Ward identities. It is subject to the Wess-Zumino consistency condition 
\bea
\label{eq_Weyl_cc}
\Delta^W_{\sigma'} \l \int d^4x \AA_\sigma^W\r - \Delta^W_\sigma\l \int d^4x \AA_{\sigma'}^W\r &=&0
\eea
(which follows from \eqref{eq_Weyl_algebra}), and is required to be scheme independent (in the sense that terms in the anomaly cannot be eliminated by addition of local functions to the generating functional). 

Again, for a discussion of the analysis of the anomaly and its implications we refer the reader to \cite{Baume:2014rla}. However, the analysis there is incomplete in two senses.
First, it is restricted to parity conserving theories, and it therefore does not involve anomalies which contain the antisymmetric tensor $\lc$.
Here we list the missing parity violating terms (still assuming that there are no chiral anomalies)\footnote{The notations for the coefficients are chosen to match the ones appearing in \cite{Baume:2014rla}.}:
\bea
\label{eq_PV_Weyl_anomaly}
\AA_\sigma^{W,\slashed P}&=&  
\sigma e ~
\epsilon^{\mu\nu\alpha\beta}R_{\mu\nu\rho\sigma}R_{\alpha\beta}^{~\rho\sigma}
\nl
&&+\sigma \epsilon^{\mu\nu\alpha\beta} \l \frac{1}{4} \tilde{\kappa}_{A B}
 F_{\mu\nu}^A F_{\alpha\beta}^B  + 
\frac{1}{2} \tilde{\zeta}_{A I J}
 F_{\mu\nu}^A \nabla_\alpha \lambda^I \nabla_\beta \lambda^J +
 \frac{1}{4} \tilde{b}_{I J K L}
 \nabla_\mu \lambda^I \nabla_\nu \lambda^J \nabla_\alpha \lambda^K \nabla_\beta \lambda^L\r \nl
\eea
where the coefficients $e,\tilde \kappa,\tilde \zeta,\tilde b$ are covariant functions of the sources $\lambda$ (antisymmetric in the $I$ indices) which are constrained by the 
following consistency conditions
\bea
\label{eq_Weyl_cc_parity_violating}
\tilde{\kappa}_{A B} P^B_I - \tilde{\zeta}_{A I J} B^J &=& 0 \nl
\tilde{\zeta}_{A I J} P^A_K - \tilde{b}_{I J K L} B^L &=& 0 
\eea
In the computation of the new consistency conditions we used the following variation rule for the field strength $ F_{\mu\nu}^A$:
\bea
\Delta^W_\sigma F^A_{\mu\nu} 
&=&\sigma\l \l f^A_{BC}S^C  - P^A_I (T_B\lambda)^I\r  F^B_{\mu\nu}- 2\d_{[J} P_{I]}^A\n_{\mu}\lambda^J\n_{\nu}\lambda^I \r 
-\n_{[\mu}\sigma\l 2  P_I^A\n_{\nu]}\lambda^I\r~. \nl
\eea
Unlike the consistency conditions discussed in \cite{Baume:2014rla}, these consistency conditions cannot be used to eliminate anomalies,
nor do they seem to imply non-trivial constraints on the RG flow. Notice that the equations \eqref{eq_Weyl_cc_parity_violating} are consistent with \eqref{eq_BP}, and lead to the following constraint on the anomaly coefficient $\wt \kappa$:
\bea
\label{eq_kappa_constraint}
\wt \kappa_{AB} P^A_I P^B_J &=& 0~.
\eea

The second element missing in the previous analyses of the Weyl anomaly was the restriction to theories with no chiral anomalies. The introduction of chiral anomalies to this framework is the subject of the next section.

 \section{Consistency of the Weyl symmetry and anomalous chiral symmetries}
 \label{sec_consistency_Weyl_chiral}

 \subsection{The Weyl anomaly and global anomalous chiral symmetries}
  \label{sec_consistency_Weyl_global_chiral}
The fact that the Weyl symmetry commutes with the flavor symmetries of the theory (see eq. \eqref{eq_Weyl_chiral_consistency}) implies the following consistency condition for the Weyl anomaly
\bea
\label{eq_Weyl_chiral_anomaly_constraint}
\Delta^W_\sigma \l \int d^4x \AA_\alpha^F\r  &=& \Delta^F_\alpha \l \int d^4x \AA_\sigma^W\r ~.
\eea
In the absence of flavor anomalies this implies that the Weyl anomaly must be a flavor singlet, however in a more general set-up this constraint has the following implications:
\begin{enumerate}
\item The consistency condition is satisfied if the chiral anomaly is Weyl invariant. 
One such Weyl invariant chiral anomaly is the chiral-gravitational anomaly \eqref{eq_chiral_gravitatinal_anomaly}\footnote{Notice that an anomaly proportional to $W^2$, the Weyl tensor squared, is not forbidden by the consistency conditions.}. 

\item In certain cases, equation \eqref{eq_Weyl_chiral_anomaly_constraint} can be used to prove the Adler-Bardeen theorem for the singlet anomaly. We can follow the logic of section \ref{sec_chiral_anomalies}, and allow for a $\lambda$ dependent coefficient in front of the anomaly 
\bea
\AA_\alpha ^F \overset{?}{=} f(\lambda)\d_\mu \alpha^A  \KK^{\mu}_A
\eea
Imposing equation \eqref{eq_Weyl_chiral_anomaly_constraint} (ignoring for a moment the Weyl variation of the background gauge fields), we find the following consistency condition
\bea
\label{eq_constraint_on_singlet_anomaly}
B^I\frac{\d}{\d\lambda^I} f(\lambda)&=&0 
\eea
We can thus conclude that the anomaly coefficient must be RG independent.
In the case where the theory has a single marginal operator, this is enough to conclude that  $f$ must be 
$\lambda$ independent, or equivalently, {the anomaly must be a 1-loop effect}.
(A similar approach is used in a proof given by Zee \cite{Zee:1972zt}). In a more general case, \eqref{eq_constraint_on_singlet_anomaly}
implies the non-trivial constraint that gradient of $f$ must be orthogonal to $B^I$.

\item
Unlike the gravitational-chiral anomaly, the $\d_\mu \KK^\mu$ anomaly \eqref{eq_chiral_anomaly} is not Weyl invariant (due to the Weyl transformation properties of the background gauge field $A_\mu^A$, see eq. \eqref{eq_Delta_W}). 
In order to make this chiral anomaly  consistent with the Weyl symmetry, we must introduce a new Weyl anomaly, whose flavor variation will match the LHS of \eqref{eq_Weyl_chiral_anomaly_constraint}.

One possibility for proceeding is to add to $\AA_\sigma^W$ the most general non-covariant terms, and then impose the consistency condition. 
Here we will use a simpler approach which is based on the following non-trivial relation between the vectors $\KK_A^\mu$ and $\XX_A^\mu$ defined in section \ref{sec_consistent_covariant}\footnote{This is a specific implementation of an equation which is used to find the covariant current in \cite{Bardeen:1984pm}.}:
\bea
\Delta_\sigma^W \l \int d^4x \d_\mu\alpha^A \KK^\mu_A\r 
&=& -
\Delta_\alpha^F \l \int d^4x  \l \delta^W_\sigma A_\mu^A\r \XX^\mu_A\r 
\eea
This relation implies that eq. \eqref{eq_Weyl_chiral_anomaly_constraint} is satisfied if the Weyl anomaly is supplemented by the following term:
\bea
\label{eq_new_weyl_anomaly}
\int d^4x \AA^W_\sigma &\supset&- \int d^4x \l \delta^W_\sigma A_\mu^A\r \XX^\mu_A=
\int d^4x  \sigma P_I^A\n _\mu \lambda ^I  \XX_A^{\mu}
+\int d^4x  \sigma S^A  \n_\mu \XX_A^{\mu}~.\nl
\eea

\end{enumerate}

This new Weyl anomaly,  in addition to insuring the consistency of the Weyl anomaly and the chiral anomaly, 
has several nice properties:
\begin{enumerate}
\item 
Writing the operator equation for $T$ \eqref{eq_T_beta_S}, keeping non-zero background gauge fields,  
we find an interesting interpretation for the new anomaly -- $T$ is given in terms of the the covariant currents defined in eq. \eqref{eq_covariant_current}, instead of the consistent ones:
\bea
[T]=
\beta^I[\OO_I]+&S^A \n_\mu [J^\mu_A] + &P_I^A \n_\mu\lambda^I [J^\mu_A]\nl
 +&S^A \n_\mu \XX^\mu_A+ &P_I^A \n_\mu\lambda^I \XX^\mu_A +\ldots \nl
=\beta^I[\OO_I]+&S^A \n_\mu [\wt J^\mu_A]  + &P_I^A \n_\mu\lambda^I [\tilde J^\mu_A] \ldots
\eea

\item 
In section \ref{sec_Weyl_flavor_no_anoamlies} we associated the ambiguity in the definition of the $\beta$-functions with the freedom to redefine the generator of the Weyl symmetry by a adding the generator of flavor transformations. In the presence of chiral anomalies eq. \eqref{eq_Weyl_ambiguity} takes the form
\bea
{\Delta_\sigma^W}' \WW \equiv \l \Delta^W_{\sigma} + \Delta^F_{\alpha=\sigma \omega}\r \WW &=&
\int d^4x \l \AA^W_\sigma + \AA^F_{\alpha=\sigma \omega} \r 
\equiv\int d^4x { \AA^W_\sigma}' 
\eea
Thanks to the specific form of the new Weyl anomaly terms, the modification of the anomaly can be absorbed by the same redefinition of the coefficient $S^A$ (see eq. \eqref{eq_operator_gauge_freedom}), plus a modification of the anomaly coefficients $\wt \kappa_{AB}$ and $e$ (see eq. \eqref{eq_PV_Weyl_anomaly})
\bea
{S^A}'&=& S^A+\omega^A\nl
{\wt \kappa_{AB}}'&=& \wt \kappa_{AB}-4d_{ABC} \omega^C\nl
{e}'&=& e-d_{A} \omega^A
\eea
where we used the relation \eqref{eq_conservation_covariant_current} and the definition of the chiral-gravitational anomaly \eqref{eq_chiral_gravitatinal_anomaly}.

In section \ref{sec_Weyl_flavor_no_anoamlies} we showed that the non-ambiguous RG flow is given in terms of the functions $B^I$ and $P_I^A$. Now we find that the non-ambiguous coefficient of the $F\tilde F$ and $R\tilde R$ Weyl anomalies are respectively 
\bea
\wt K_{AB}&\equiv& \wt \kappa_{AB}+4d_{ABC} S^C\nl
E&\equiv&e+d_AS^A
\eea

Given the functions $B$ and $P$ we factorized out from the generator of the Weyl symmetry a flavor rotation, controlled by the parameter 
$S$. Schematically, we showed that
\bea
\Delta_\sigma ^W(\beta,\rho,S) &=& \Delta_\sigma ^W( B, P,0 ) + \Delta_{\sigma S} ^F
\eea
Using the new terms found in \eqref{eq_new_weyl_anomaly}, the same can be done for the anomaly:
\bea
\label{eq_anomaly_S}
\AA_\sigma^W (\kappa, e,S) = \AA^W_\sigma (K,E,0)+ \AA_{\sigma S}^F~.
\eea
We conclude that even in the presence of chiral anomalies, it is still possible to consistently decompose the local RG equation into a non-ambiguous component plus an anomalous flavor Ward identity.

\item The new anomaly must satisfy the Weyl Wess-Zumino consistency condition \eqref{eq_Weyl_cc}. An explicit computation shows that the cancelation of the contribution of this new anomaly to the RHS of \eqref{eq_Weyl_cc} is achieved simply by replacing the coefficient $\wt \kappa_{AB}$ in the consistency condition \eqref{eq_Weyl_cc_parity_violating} with the non-ambiguous function $\wt K_{AB}$. 
We conclude that, as should be expected, the consistency conditions are written in terms of functions which are independent of the ambiguity related to the wavefunction renormalization.

\end{enumerate}

 \subsection{The Weyl symmetry and the $\theta$ parameter}
 \label{sec_running_theta}

In section \ref{sec_theta} we introduced the $\theta$ parameter as a compensator for anomalies involving dynamical gauge fields. The symmetry generator was given by \eqref{eq_axial_generator}.
In the presence of the $\theta$ background field the generator of Weyl anomalies should be generalized as follows:
\bea
\Delta_\sigma^W&=&\int d^4x \l \sigma 
 \l 2 g^{\mu\nu}\frac{\delta}{\delta g^{\mu\nu}(x)}
-\beta^I\frac{\delta}{\delta \lambda^I(x)}
-\beta^\theta\frac{\delta}{\delta \theta(x)}\r 
-\l \sigma \rho^A_I\n_\mu\lambda^I
- \d_\mu\sigma S^A\r \frac{\delta }{\delta A_\mu^A(x)}
\r \nl
\eea
Notice that none of the dimensionless $\beta$-functions (as well as the anomaly coefficients) can depend explicitly on $\theta$ 
due to the constraint that $\theta$ appears only with a space-time derivative, however the summation over the sources in the $\rho$ term contains a derivative of $\theta$ as well ($\rho^A_I\n_\mu\lambda^I \supset  \rho^A_\theta\n_\mu\theta$. Similarly, the Weyl anomaly may contain gradients of $\theta$). The index $A$ runs over the symmetry generators of the theory, and we denote the generator the axial symmetry as $T_A=T_5$.

Next, we extract the Ward identity with parameter $\sigma S$ as was done in \eqref{eq_extracting_flavor_rotation}:
\bea
\label{eq_Delta_W_B_theta}
\Delta_\sigma^W&=&\int d^4x \sigma 
\l 2 g^{\mu\nu}\frac{\delta}{\delta g^{\mu\nu}(x)}
-B^I\frac{\delta}{\delta \lambda^I(x)}
-B^\theta\frac{\delta}{\delta \theta(x)}
- B_\mu^A \frac{\delta }{\delta A_\mu^A(x)}
\r -\Delta^F_{\sigma S}\nl
\eea
where
\bea
B^\theta&=& \beta^\theta +  S^5
\eea
and $S^5$ is the component of $S^A$ associated with the anomalous axial symmetry.

Let us comment on the renormalization of the anomaly operator, $\d_\mu \hat\KK^\mu_5$, sourced by $\theta$. As explained in \cite{Baume:2014rla},
the RG variation of a nearly-marginal operators is determined by the formula
\bea
\left [\Delta^{RG}, \frac{\delta}{\delta \lambda^I}\right ]&=&\d_I B^J\frac{\delta}{\delta \lambda^I} + P_I^A \n_\mu \frac{\delta}{\delta A_\mu^A}
\eea
Applying this formula to the source $\theta$, and using the fact that $\d_\theta B^J=0$, we find that the anomaly 
operator $\d_\mu \hat\KK^\mu_5$ is renormalized only by divergences of currents, as was discussed in section \ref{sec_theta} (see also \cite{Larin:1993tq}).

Another implication of the $\theta$-independence of the coefficients in the local RG equation is a relation between $B^\theta$ and the remaining $\beta$-functions of the theory. This relation is based on the gradient flow formula, a formula which was derived in \cite{Osborn:1991gm} using the Weyl WZ consistency conditions, and takes the general form
\bea
\frac{\d}{\d \lambda^I} \tilde a &=& \chi_{IJ}B^J
\eea
where $\tilde a$ and $\chi_{IJ}$ are combinations of various coefficients in the Weyl anomaly.
 The form of the equation is unchanged if the indices $I$ or $J$ correspond to the source $\theta$ \footnote{
The Weyl variation of $\n_\mu\theta$ defined in \eqref{eq_covariant_derivatives} is given by 
\bea
\Delta_\sigma ^W  (\n_\mu\theta)
&=&-\d_\mu\sigma B^\theta-\sigma \n_\mu \lambda^I \l \d_I B^\theta - P^5_I\r 
\eea
This implies that the analysis of Weyl variation of functions of the sources, which is used in the derivation of the gradient flow formula as described in \cite{Baume:2014rla}, is basically unchanged in the presence of the $\theta$ field, and the gradient flow formula can be generalized to include $\theta$. 
}.
Using the fact that the function $\tilde a$ must be independent of $\theta$, we find the following non-trivial relation
\bea
\label{eq_theta_gradient_flow}
\chi_{\theta\theta}B^\theta &=& - \chi_{\theta I}B^I~.
\eea
Assuming  $\chi_{\theta\theta}$ is non-zero\footnote{The Weyl anomaly for a Yang-Mills theory in the presence of an $x$-dependent $\theta$ was computed in \cite{Osborn:2003vk} and $\chi_{\theta\theta}$ was found to be non-zero.}, this equation rules out the possibility for $B^\theta$ to be the only non-vanishing $\beta$-function. Indeed, such a scenario would correspond to an exotic RG flow where only the $\theta$ parameter is running. A theory with this property is not a conformal theory, yet it is invariant under global rescaling of the metric because $\frac{\delta}{\delta \theta}\WW$ is a total derivative (in perturbation theory).
It is interesting how the Weyl consistency conditions can be used to rule out this scenario.

It is a standard procedure to eliminate the $\theta$ parameter by an axial rotation and absorb it into the phase of the fermion mass matrix or Yukawa couplings. 
In order to make sure that it is not generated by the RG flow, one has to eliminate the running of the $\theta$ parameter.
This can be done by rewriting eq. \eqref{eq_Delta_W_B_theta} in the following form
\bea
\Delta_\sigma^W&=&\int d^4x \l \sigma 
\l 2 g^{\mu\nu}\frac{\delta}{\delta g^{\mu\nu}(x)}
-B^{I,\theta}\frac{\delta}{\delta \lambda^I(x)}
- B_\mu^{A,\theta} \frac{\delta }{\delta A_\mu^A(x)}
\r
 -\d_\mu\sigma B^\theta\frac{\delta }{\delta A_\mu^5(x)}\r 
-\Delta^F_{\sigma S}-\Delta^5_{\sigma B^{\theta}}\nl
\eea
where
\bea
B^{I,\theta}&=&B^I +  B^\theta(T_5 \lambda)^I\nl
B_\mu^{A,\theta} &=&B_\mu^A +\delta^A_5\n_\mu B^\theta  
\eea
We see that in the scheme where $\theta$ is not running, the $\beta$ functions are modified by a term proportional to $B^\theta$.
Furthermore, using eq. \eqref{eq_theta_gradient_flow} (assuming $\chi_{\theta\theta}$ is non-zero) the $\beta$ functions in this scheme can be expressed as
\bea
B^{I,\theta}&=&B^J\l \delta^I_J -\frac{\chi_{\theta J}}{\chi_{\theta\theta}} (T_5 \lambda)^I\r~.
\eea

 \section{ Discussion}
\label{sec_discussion}

The background source method and the local RG equation are efficient tools for studying RG flows in a model independent way. The assumption that the generating functional $\WW$ is renormalized allows us to bypass difficulties associated with the regularization of the theory, and the symmetry generators $\Delta_\alpha^F$ and $\Delta_\sigma^W$ provide a compact formalism for generating Ward identities and deriving consistency conditions.

In this work we added two ingredients to this framework -- we found the consistency conditions for parity violating Weyl anomalies (eq. \eqref{eq_Weyl_cc_parity_violating}) and we verified that the local RG equation is consistent with chiral anomalies. The main results are the new Weyl anomaly \eqref{eq_new_weyl_anomaly} and the fact, which is implied by eq. \eqref{eq_anomaly_S}, that the $S$ dependent flavor rotations can be factored out and eliminated by a choice of scheme, even in the presence of chiral anomalies.

As is demonstrated in section \ref{sec_sources_and_flavor}, the background source method is useful in analyzing chiral anomalies even without imposing the Weyl symmetry. It reflects the non-renormalization of Ward identities and can be used to prove the Adler-Bardeen theorem for non-abelian theories as a direct consequence of the WZ consistency conditions.

In addition, we found new constraints on the RG flow in the presence of the $\theta$ parameter. The results described in sec. \ref{sec_running_theta} can be used to study the running of CP violating parameters in the standard model and possible solutions to the strong CP problem (see e.g. \cite{Ellis:1978hq},\cite{Nelson:1983zb}). We leave this possibility for future research.

\section*{Acknowledgments}

We would like to thank Y. Burnier, C. Hoyos, J. Fortin, A. Monin, R. Rattazzi, H. Osborn and L. Vitale for useful discussions.
This work was supported by the Swiss National Science Foundation under grants 200020-138131 and 200020-150060.

\section*{Appendix}
\addtocontents{toc}{\protect\setcounter{tocdepth}{1}}

\appendix

\section{WZ consistency conditions in the presence of background sources}
\label{app_WZ_CC_BG_sources}
In this appendix we classify the possible chiral anomalies in the presence of background sources, checking whether they satisfy the Wess-Zumino consistency conditions, and whether they can be written as variations of local functions. For this purpose we consider all the possible combinations of Lorentz scalars constructed from the background gauge fields $A_\mu^A$ and a generic covariant function of the sources which we denote by $K$ (For example, a possible candidate for scalar function $K$ of dimension 4 is $K=\sqrt {-g}f(\lambda)^A_{IJ} T_A G^{\mu\nu}\n_\mu\lambda^I\n_\nu\lambda^J$ where $G^{\mu\nu}$ is the Einstein tensor associated with the background metric and $\n_\mu$ is a covariant derivative).

\subsection{Non-Abelian symmetries}
In the case of non-abelian anomalies, we assume that $K$ transforms as $\Delta^F_\beta K= [K,\beta]$
and compute for each possible anomaly term the violation of the WZ condition, namely 
\bea
\delta^{WZ}_{\alpha\beta}
&\equiv&\Delta^F_\beta \l \int d^4x \AA^F_\alpha \r - \Delta^F_\alpha \l \int d^4x \AA^F_\beta \r
-\int d^4x \AA^F_{ [\beta,\alpha]}~. 
\eea
\begin{enumerate}
\item
First, let us consider terms which cannot be written as total derivatives:

\begin{tabular}{ l l  }
   $\AA_\alpha$ &$   \delta^{WZ}_{\alpha\beta}$ \tabspace   \\  \hline\hline
  $\Tr {\alpha K}$ & $2\Tr {[\beta,\alpha] K}$  \\
  $\Tr {\alpha K^\mu A_\mu}$ & $2\Tr {[\beta,\alpha] K^\mu A_\mu } +  \Tr{\alpha K^\mu \d_\mu \beta }- \Tr{\beta K^\mu \d_\mu \alpha }$  \tabspace   \\ 
  $\Tr {\alpha K^{\mu\nu\ldots} A_\mu A_\nu\ldots}$ & $2\Tr {[\beta,\alpha] K^{\mu\nu\ldots} A_\mu A_\nu\ldots }
   +  \Tr{\alpha K^{\mu\nu\ldots} \d_\mu \beta A_\nu\ldots }
    -\Tr{\beta K^{\mu\nu\ldots} \d_\mu \alpha A_\nu\ldots }+\ldots
$  \tabspace   \\ \hline
\end{tabular}

There are no anomalies whose contribution to $\delta^{WZ}_{\alpha\beta}$ can cancel these terms. We conclude that a non-abelian chiral anomaly must be written as a total-derivative. 

\item 
The next family of possible terms are either inconsistent, or can be written as variation of local terms

\begin{tabular}{ l l l }
   $\AA_\alpha$ &$\delta^{WZ}_{\alpha\beta}$ &  comments\tabspace   \\  \hline\hline
  $\Tr {\d_\mu \alpha K^\mu}$ & $0$ & $ =\Delta^F_\alpha \l \int d^4x\Tr {A_\mu K^\mu}\r $  \tabspace  \\
  $\Tr {\d_\mu \alpha \l K^{(\mu\nu)} A_\nu + A_\nu K^{(\mu\nu)}\r  }$ & $0$ & $  =\Delta^F_\alpha \l \int d^4x \Tr {A_\mu K^{(\mu\nu)}A_\nu}\r $\tabspace \\ 
$\Tr {\d_\mu \alpha \l K^{[\mu\nu]} A_\nu - A_\nu K^{(\mu\nu)}\r  }$ & $0$ & $
  =\Delta^F_\alpha \l \int d^4x\Tr {A_\mu K^{[\mu\nu]}A_\nu}\r $  \tabspace   \\
  $\Tr {\d_\mu \alpha \l K^{(\mu\nu)} A_\nu - A_\nu K^{(\mu\nu)}\r  }$ & $2\Tr{ K^{(\mu\nu)}(\d_\nu\beta\d_\mu \alpha - \d_\nu\alpha\d_\mu \beta)}
 $\\
  $\Tr {\d_\mu \alpha \l K^{[\mu\nu]} A_\nu + A_\nu K^{[\mu\nu]}\r  }$ & $2\Tr{ K^{[\mu\nu]}(\d_\nu\beta\d_\mu \alpha - \d_\nu\alpha\d_\mu \beta)}  
  $ & \tabspace  \\  \hline
\end{tabular}

($(\mu\nu)$ and $[\mu\nu]$ denote symmetric and anti-symmetric tensors, respectively). Notice that the last term can be eliminated by integration by parts only if $K^{[\mu\nu]}$ satisfies $\d_\mu K^{[\mu\nu]} = 0$.

\item
Terms with two powers of the background gauge field can be parameterized by

\begin{tabular}{ l l l }
   $\AA_\alpha$ &$\delta^{WZ}_{\alpha\beta}$ &  comments\tabspace   \\  \hline\hline
$\lc \Tr {\d_\mu \alpha  K_\nu A_\rho A_\sigma }$&$\lc \Tr{
 K_\nu\d_\rho \beta  A_\sigma\d_\mu \alpha  - K_\nu\d_\rho \alpha  A_\sigma \d_\mu \beta 
} $&\tabspace \\
&$+\lc \Tr{
  K_\nu A_\rho (\d_\sigma \beta\d_\mu \alpha-\d_\sigma \alpha\d_\mu \beta)
 } $&\tabspace \\
$
 \lc \Tr {\d_\mu \alpha  A_\rho K_\nu  A_\sigma } $&$
 \lc \Tr{
(\d_\mu \alpha \d_\rho \beta -\d_\mu \beta \d_\rho \alpha) (K_\nu  A_\sigma - A_\sigma K_\nu )} $&\tabspace \\
$
\lc \Tr {\d_\mu \alpha  A_\rho A_\sigma K_\nu }$&$
 \lc \Tr{K_\nu \d_\mu \alpha  A_\rho \d_\sigma \beta 
 -K_\nu \d_\mu \beta  A_\rho \d_\sigma \alpha  }$& \tabspace \\ 
&$
+ \lc \Tr{
(\d_\mu \alpha \d_\rho \beta -\d_\mu \beta \d_\rho \alpha ) A_\sigma K_\nu  }$& \tabspace \\ 
Sum of the above
&$0$& 
$ = \Delta^F_\alpha \l \Tr {A_\mu A_\nu A_\rho K_\sigma}\r $ \tabspace \\ 
\hline
\end{tabular}

We see that the only consistent combination can be eliminated by a choice of scheme.

\item
The terms appearing in the consistent anomaly are the following:

\begin{tabular}{ l l  }
   $\AA_\alpha$ &$\delta^{WZ}_{\alpha\beta}$ \tabspace   \\  \hline\hline
$\lc \Tr {\d_\mu \alpha  A_\nu A_\rho A_\sigma }$&$
2 \lc \Tr{ (\d_\nu\beta\d_\mu \alpha - \d_\nu\alpha\d_\mu \beta)A_\rho A_\sigma}
 $\tabspace \\
$\lc \Tr {\d_\mu \alpha  \l A_\nu F_{\rho\sigma} +  F_{\rho\sigma} A_\nu \r }$&$
2 \lc \Tr{ (\d_\nu\beta\d_\mu \alpha - \d_\nu\alpha\d_\mu \beta)A_\rho A_\sigma}
 $\tabspace \\
\hline
\end{tabular}

The linear combination \eqref{eq_chiral_anomaly},\eqref{eq_chiral_anomaly_K} is therefore consistent.

\item
Allowing for an arbitrary, flavor singlet, coefficient $K(\lambda)$ in front of the consistent anomaly, we find an obstruction for satisfying the consistency condition:

\begin{tabular}{ l l }
   $\AA_\alpha$ &$\delta^{WZ}_{\alpha\beta}$  \tabspace   \\  \hline\hline
$K\lc \Tr {\d_\mu \alpha  A_\nu A_\rho A_\sigma }$&$
2K \lc \Tr{ (\d_\nu\beta\d_\mu \alpha - \d_\nu\alpha\d_\mu \beta)A_\rho A_\sigma}
 $\tabspace \\
&$
-2 \lc \Tr{ (\d_\nu\beta\d_\mu \alpha - \d_\nu\alpha\d_\mu \beta)  A_{\sigma}}\d_\rho K 
 $\tabspace \\
$K\lc \Tr {\d_\mu \alpha  \l A_\nu F_{\rho\sigma} +  F_{\rho\sigma} A_\nu \r }$&$
2K \lc \Tr{ (\d_\nu\beta\d_\mu \alpha - \d_\nu\alpha\d_\mu \beta)A_\rho A_\sigma}
 $\tabspace \\\hline
\end{tabular}

This is a quick derivation of the Adler-Bardeen theorem for the case of non-abelian anomalies.
A similar conclusion is reached when considering non-flavor-singlet functions $K$.
\end{enumerate}

\subsection{Singlet anomalies}

The consistent anomaly of an abelian symmetry must satisfy the constraints described for the non-abelian case,
with the following exception -- a flavor singlet $\AA^{U(1)}_\alpha=\alpha K$ is a consistent anomaly.
However, such anomalies are constrained by the consistency with respect to the Weyl symmetry  
\bea
\label{eq_singlet_anomaly_weyl_cc}
\Delta^W _\sigma \l \int d^4x \AA_\alpha^{U(1)}\r &=& \Delta^{U(1)} _\alpha\l \int d^4x  \AA_\sigma^W\r 
\eea
In addition to the interesting structure which satisfies this constraint and is discussed in section \ref{sec_consistency_Weyl_global_chiral}, another possibility are singlet, Weyl invariant terms which automatically have vanishing contributions to the LHS of eq. \eqref{eq_singlet_anomaly_weyl_cc}, and do not require introduction of new Weyl anomalies.
One such term is the mixed chiral-gravitational anomaly
\bea
\AA_\alpha^{U(1)}&\propto &\epsilon^{\mu\nu\alpha\beta}R_{\mu\nu\rho\sigma}R_{\alpha\beta}^{~\rho\sigma}
\eea
Interestingly, the consistency conditions allow for a chiral anomaly proportional to the Weyl tensor squared. This is the only new candidate for a consistent anomaly we found.


\end{document}